\documentclass[pra,twocolumn,superscriptaddress,floatfix]{revtex4}
\usepackage{amsfonts}
\usepackage{amsmath}
\usepackage{amssymb}
\usepackage[sort&compress]{natbib} 
\usepackage{graphicx, times}
\usepackage{color}
	
\begin{document}

\title{Stable and metastable freezing of classical correlations in qutrits}

\author{C.E. L\'opez}

\affiliation{Departamento de Física, Universidad de Santiago de Chile, USACH,
Casilla 307 Correo 2 Santiago, Chile}

\author{F. Lastra}

\affiliation{Departamento de F\'isica, Facultad de Ciencias B\'asicas, Universidad
de Antofagasta, Casilla 170, Antofagasta, Chile}

\date{\today}
\begin{abstract}
We study the dynamics of quantum and classical correlations in a two-qutrit system coupled to independent reservoirs. In particular, we addressed the differences in the dynamics of Markovian and non-Markovian regimes and show that for specific initial states, classical correlations exhibit abrupt changes along the dynamics. A particular sudden change occurs when the classical correlations freezes to a certain value at a given time, revealing the apparition of a pointer-state basis.   After this given time, the decoherence only affects quantum correlations. Here we identify two regimes in the decoherence dynamics: a mixed regime when both classical and quantum correlations decay and a quantum regime when only quantum correlations decay.  We show that the freezing of classical correlations can be stable or metastable depending on the system-reservoirs parameters. In the long-time limit, we find analytical expressions for the pointer-state basis the system settles in, and consequently for classical and quantum correlations.
\end{abstract}

\pacs{pacs}

\maketitle

\section{introduction}
Quantum to classical transition has been an interesting subject since the beginning of quantum theory \cite{Joos}. This transition can be described as a flow of correlations from the quantum system to its surroundings \cite{Breuer,Zurek00,Zurek03}. Because of that, all correlations that can be shared by two parts can not be preserved and begin to be lost as time goes by. Recently, to characterize when a quantum system has began to lose its quantum component has regained the attention of researchers \cite{Mazz,Xu,Cor,Pau,Las}. On the other hand, it is well known the complexity in the maximization procedure to obtain any correlation shared by the subsystems when the dimension of the Hilbert space is increased more than 2 in each of them. This is a hard task, indeed, there exist a few cases where analytical expressions have been found, to name the most relevant we have: entanglement in $2\otimes 2$ dimensional  systems \cite{Woo} and some families of states \cite{Ter,Hor}, Classical and quantum correlations \cite{Luo,Xstate,Karpat,Chi,Ali,Chen,Khan,Khan1}. If we consider the total quantum system made of two parts, then, all correlations both classical and quantum can be defined as entropic quantities \cite{Olli01}. In what follows, we will define the total correlations by means of measures on one of the subsystems.

A bipartite quantum system $\hat{\rho}_{{\rm AB}}$ can feature both quantum and classical correlations. Total correlations can be characterized by the quantum mutual information \cite{Olli01,Hen,Ved,Gro,Schu}

\begin{equation}
I(\hat{\rho}_{{\rm AB}})=S(\hat{\rho}_{{\rm A}})+S(\hat{\rho}_{{\rm B}})-S(\hat{\rho}_{{\rm AB}}),\label{qmi}
\end{equation}
where $S(\hat{\rho})=-{\rm Tr}[\hat{\rho}\lg(\hat{\rho})]$ is the
von Neumann entropy. Based on this expression
it is commonly believed that the correlations can
be separated according to their classical and quantum
nature, respectively \cite{Olli01}. In this way the quantum discord
has been introduced as \cite{Ollivier}

\begin{equation}
D(\hat{\rho}_{AB})=I(\hat{\rho}_{AB})-C(\hat{\rho}_{AB})\label{qd}
\end{equation}
where $C(\hat{\rho}_{AB})$ are the classical correlations \cite{Olli01, Hen} defined
by the following maximization procedure: A complete set of projector
operators $\{\hat{\Pi}_{k}\}$ must be constructed for the subsystem
$B$. Then the quantity 
\begin{equation}
C(\hat{\rho}_{AB})=\max_{\{\hat{\Pi}_{k}\}}\left[S(\hat{\rho}_{A})-S(\hat{\rho}_{AB}\mid\{\hat{\Pi}_{k}\})\right],\label{cc}
\end{equation}
must be maximized with respect to variation of the set of $\{\hat{\Pi}_{k}\}$
where $S(\hat{\rho}_{AB}\mid\{\hat{\Pi}_{k}\})=\sum_{k}p_{k}S(\hat{\rho}_{k})$,
$p_{k}={\rm Tr}(\hat{\rho}_{AB}\hat{\Pi}_{k})$, and $\hat{\rho}_{k}={\rm Tr}_{B}(\hat{\Pi}_{k}\hat{\rho}_{AB}\hat{\Pi}_{k})/p_{k}$.

In this paper, we investigate the dynamical evolution of classical correlations, using models for decoherence in the Markovian as well as non-Markovian regime. Our main focus is to study the evolution of classical and quantum correlations in the case of a bipartite system, where each part is represented by a subsystem of dimension three or so called qutrit. In Section II, we begin by introducing a general decoherence model from which the two regimes mentioned above can be obtained. These regimes can be implemented simply by making assumptions on the time given by the inverse of decay rates and correlation time of the reservoir. Also, we define the initial state for two qutrits using the discrete quantum fourier transform. In Section III, We present a general base in the qutrit Hilbert space where the maximization procedure can be carried out. Although its form is simple, it is worth to mention that in the actual case maximization procedure must be performed in an bloch hipersphere defined by four angles. Following in this section, we specialize our analysis studying three cases of interest: A. The Markovian case, B. The non-Markovian case and C. The limit of long times. We present analytic results for classical correlations, showing the apparition of stable and metastable pointer states. Finally, in Section IV, we present our concluding remarks.        

\section{Quantum dynamics}
In this manuscript, we will consider the dynamics of a bipartite qutrits system under the onset of dephasing. In this scenario, the evolution of the system is governed by the master equation:

\begin{eqnarray}
\dot{\hat{\rho}}_{AB} &=&\sum_{j=A,B}\frac{Q_{j}(t)}{2}[2 \hat{f}_j ^ {\dagger } \hat{f}_j \hat{\rho}_{AB} \hat{f}^{\dagger }\hat{f}_j-\left(
\hat{f}_{j}^{\dagger }\hat{f}_j\right) ^{2}\hat{\rho}_{AB} \notag \\
&& -\hat{\rho}_{AB} \left( \hat{f}_j^{\dagger }\hat{f}_j\right) ^{2}] \label{master}
\end{eqnarray}
where $\hat{f}_j$ are operators acting on the $j-$th system. This master equation allows us to study two different regimes: Markovian and non-Markovian. To do this we consider that the reservoirs present Ornstein-Uhlenbeck correlations, where \cite{eberly2012}

\begin{eqnarray*}
Q_j (t)&=&\frac{\Gamma_j \gamma_j}{2}\left[\frac{\sin(\eta_j t)}{\eta_j \cos(\eta_j t)+(\gamma_j /2) \sin(\eta_j t)}\right] 
\label{qnom},
\end{eqnarray*}
where $1/\gamma_i$ is the correlation time of the reservoirs, $\Gamma_i$ is the decay rate of the qutrits subsystem and $\eta_i^2=(\Gamma_i - \gamma_i /2)\gamma_i /2$. Solving the master equation (\ref{master}), we find that in the basis: $\{|n m\rangle\}$ with $n,m=0,1,2,\dots d-1$, with $d$ the dimension of each system, the density matrix elements are given  by
\begin{equation}
\rho _{nk,ml}(t)=\rho _{nk,ml}(0) P_A(t)^{\left\vert n-m\right\vert ^{2}}P_B(t)^{\left\vert k-l\right\vert
^{2}}
\label{sol1}
\end{equation}
where, 
\begin{equation}
P_j(t)=e^{\beta_j t}\left( \cos{(\eta_j t)}-\frac{\beta_j} {\eta_j} \sin{(\eta_j t)}\right)
\label{p}
\end{equation}
with $\beta_j =-\gamma_j / 2 $.

In order to study the evolution of quantum and classical correlations in higher dimensional bipartite systems, the election of the initial state must ensure that the system is actually occupying  more than two dimensions of the Hilbert space. Here, we consider an incoherent superposition of generalized Bell states for qutrits as follows.						
\begin{equation}
\hat{\rho}_{AB}(0)=p_0 |\phi_{00}\rangle \langle\phi_{00}| +p_1 |\phi_{01}\rangle \langle \phi_{01}|+p_2 |\phi_{02}\rangle \langle \phi_{02}|\label{eq:qubit-state}
\end{equation}
The generalized Bell states are defined as
\begin{equation}
|\phi_{jk}\rangle = \hat{X}_{12} \hat{F}_1 |jk\rangle_{12} \label{transf}	
\end{equation}
where $\hat{X}_{12}$ is the XOR-gate and is defined through $ \hat{X}_{12} |j\rangle|k\rangle= |j\rangle|j \ominus k\rangle$ with  $j \ominus k$ is the difference between $j$ and $k$ modulus $d$, with $d$ being the dimension of each system. The operator $\hat{F}$ is the discrete quantum Fourier transform and is defined acting on the state $|j\rangle$, leading to $\hat{F} |j\rangle = (1/\sqrt{d})\sum_{k=0}^{d-1} \exp{(i 2\pi jk/d) }|k\rangle$. Notice that for $d=2$, $\hat{X}_{12}$ is the controlled NOT gate, and the Fourier transform is the Hadamard gate. This two operators acting on the two-qubit basis $\{|00\rangle,|01\rangle,|10\rangle,|11\rangle\}$, generate all four Bells states for two-qubit systems.  Now for qutrits ($d=3$), from Eq.~(\ref{transf}) we have that
\begin{eqnarray*}
|\phi_{00}\rangle & = & \frac{1}{\sqrt{3}} \left(|00\rangle+ |11\rangle+ |22\rangle \right)\\
|\phi_{01}\rangle & = & \frac{1}{\sqrt{3}} \left(|02\rangle+ |10\rangle+ |21\rangle \right)\\
|\phi_{02}\rangle & = & \frac{1}{\sqrt{3}} \left(|01\rangle+ |12\rangle+ |20\rangle \right)
\end{eqnarray*}

\section{Quantum and classical correlations dynamics}

To calculate the evolution of quantum and classical correlations in our qutrit system, we have to choose a general set of three orthogonal states. This set must be constructed in  such a way that the measurement projectors cover the complete Bloch sphere. Furthermore, as measurements can be performed on either qutrits we must choose one of them. Here for instance, classical correlations will be calculated by performing measurements on qutrit $B$. For that, we consider the basis \cite{Caves}:
\begin{eqnarray}
|V_{1}\rangle &=& e^{i\chi_1}\sin\theta\cos\phi|0\rangle+e^{i\chi_2}\sin\theta\sin\phi|1\rangle+\cos\theta|2\rangle\notag \\
|V_{2}\rangle &=& e^{i\chi_1}\cos\theta\cos\phi|0\rangle+e^{i\chi_2}\cos\theta\sin\phi|1\rangle-\sin\theta|2\rangle  \notag\\
|V_{3}\rangle &=&-e^{i\chi_1}\sin\phi|0\rangle+e^{i\chi_2}\cos\phi|1\rangle \notag,
\end{eqnarray}
where the range for the angles are $0 \le \theta,\phi \le \pi/2$ and $0 \le \chi_1,\chi_2 \le 2\pi$. Using this basis we can evaluate the expression for classical correlations given in Eq.~{(\ref{cc}). Although has been argued that more general measurement should be considered to calculate classical correlations, it has been shown that this generates only minimal corrections to the calculations using projective measurements~\cite{Galve}. 

In the following, we will study the quantum and classical correlations in both Markovian and non-Markovian regimes. 

\subsection{Markovian Regime}

The Markovian regime is recovered when the reservoir correlation time becomes much smaller than the system decay time $(\Gamma_j \ll \gamma_j)$. In this limit, it can be shown that $P(t)_j \approx e^{-\frac{1}{2}\Gamma_j t}$ and the density matrix elements of Eq.~(\ref{sol1}) reduce to
\begin{equation}
\rho _{nk,ml}(t)=\rho _{nk,ml}(0)\exp \left[ -\frac{1}{2}(\Gamma
_{1}\left\vert n-m\right\vert ^{2}+\Gamma _{2}\left\vert k-l\right\vert
^{2})t\right].
\label{sol2}
\end{equation}

The evolution of quantum and classical correlations for this density matrix are shown in Fig.~\ref{fig1markov} for the initial state of Eq.~(\ref{eq:qubit-state}), for four different set of parameters $p_0$, $p_1$ and $p_2$. Fig.~\ref{fig1markov} (a) corresponds to the case with different values of the parameters ($p_0=0.3, p_1 = 0.1$ and $p_2 = 0.6$). Interestingly we observe in this case that, until a given (finite) time $\Gamma t$, the classical correlations decay. Then, it freezes to an stationary value while quantum discord decays asymptotically to zero. In previous works \cite{Mazz,Cor,Las}, a similar behavior was found in the two-qubit scenario where the classical correlation also exhibits a sudden change in its dynamics accompanied by a sudden change in the discord dynamics. However, this is not longer true in our case since the quantum discord decay at all times. This can be interpreted as the decoherence dynamics exhibiting two regimes: a mixed one, where decoherence has a quantum and a classical contribution, i. e., both correlations decay; and a second regime where the decoherence has only a quantum character. In this last regime, only the quantum correlations decay.  

Although the evolution of correlations is similar, in Fig.~\ref{fig1markov}(b), we observe that for a different set of parameters ($p_0=0, p_1 = 0.5$ and $p_2 = 0.5$) the stationary value of the classical correlations is considerably higher compared to the case $(a)$. This becomes more apparent when we consider an initial state with parameters $p_0 = p_1 = 0$ and $p_2 =1$. This particular state corresponds to a pure initial state whose quantum and classical correlations evolve as shown in Fig.~\ref{fig1markov}(c). Interestingly, the classical correlations for this state are not affected by decoherence and stay constant along the dynamics, while the quantum discord decays asymptotically to zero as expected. 

The sudden change in the classical correlations depicted in Fig.~\ref{fig1markov}(a) and~\ref{fig1markov}(b), reveals the apparition of a pointer state associated to the system being measured as have been encountered in previous works \cite{Cor,Las}. The stationary value reached by the classical correlations tells us that, by measuring on the system $B$, we will obtain the same information about the $A$ at all times and, the measurement operators are defined in basis of classical states that are not affected by decoherence \cite{Cor}.  In other words, we observe that after a finite time, the system settles on an stable pointer state basis. 

On the other hand, in Fig.~\ref{fig1markov}(d) we show the case with $p_0 =p_1 =p_2 = 1/3$. Interestingly, for equal parameters in the initial state, the classical correlations show a different behavior compared to the previous ones: it decays asymptotically to zero as well as the quantum discord. Thats is, for a balanced incoherent superposition in the initial state~(\ref{eq:qubit-state}), the systems does not reach a pointer state. Numerically we have observed that whichever the combination of $p_j$'s is, we always found a non-zero stationary value for classical correlations with the only exception of  $p_0 =p_1 =p_2 = 1/3.$ 

\begin{figure}
\includegraphics[width=80mm]{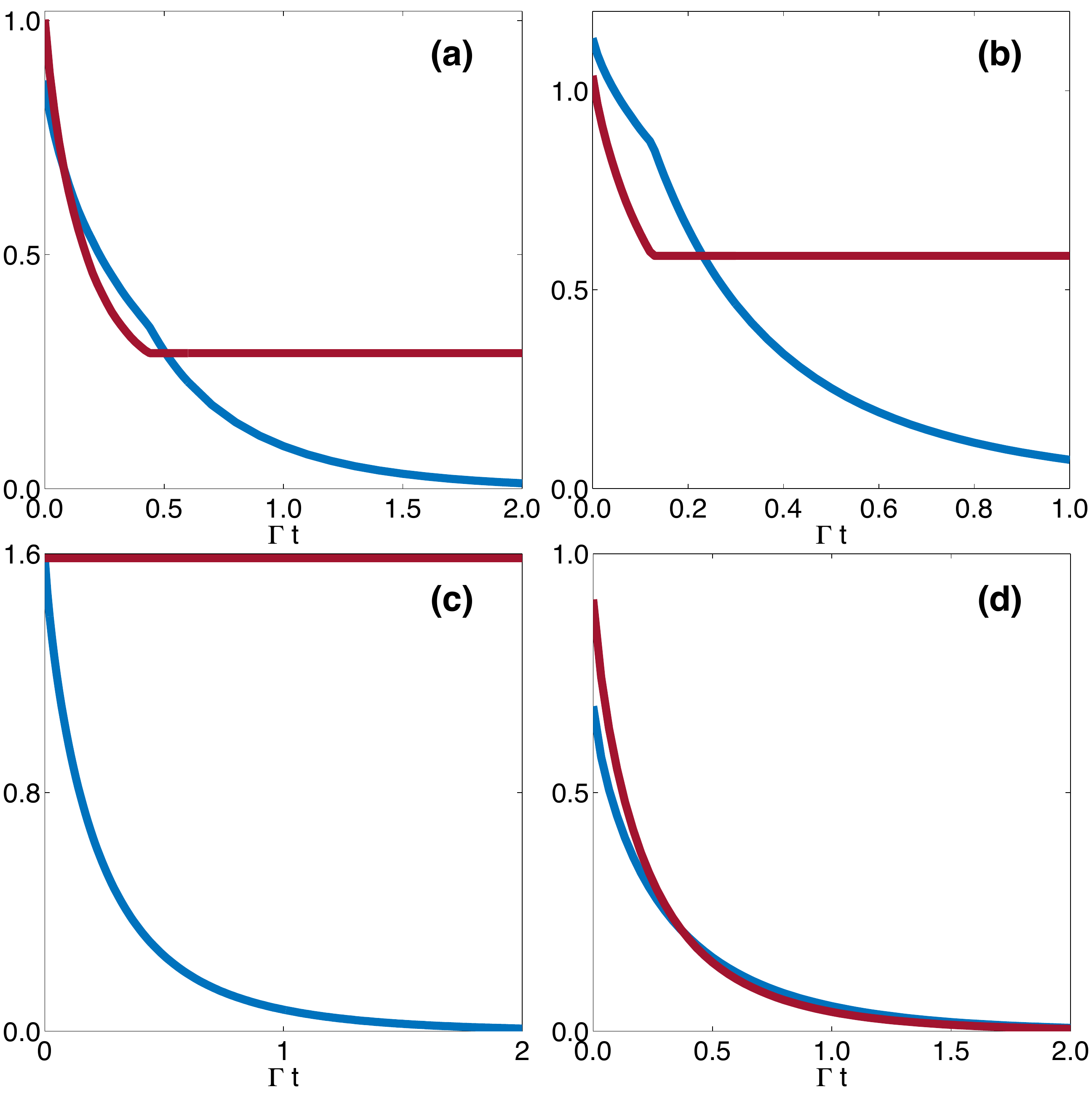} \caption{Evolution of  discord $D(\hat{\rho}_{AB})$ (blue) and classical correlations $C(\hat{\rho}_{AB})$ (red) for the initial state of Eq.~(\ref{eq:qubit-state})  with parameters $(p_0, p_1, p_2)$ with values  (a): $(0.3, 0.1, 0.6)$, (b) $(0, 0.5, 0.5)$, (c): $(0,0,1)$ and (d): $(1/3,1/3,1/3)$. For simplicity we consider $\Gamma_1 = \Gamma_2 = \Gamma$.}
\label{fig1markov} 
\end{figure}

\subsection{Non-Markovian Regime}
In this regime we will consider the cases when $\Gamma \sim \gamma$ and $\Gamma \gg \gamma$. This relation between coherence times for reservoirs and qutrits allows the quantum system to exhibit a richer dynamics and in consequence, quantum and classical correlations show features that can not be observed under de Markov approximation.

For example, in Fig.~\ref{fignonmarkov}  we show the evolution of quantum and classical correlations considering the initial state (\ref{eq:qubit-state})  with $p_0 = 0.3$, $p_1 =0.1$ and $p_2 = 0.6$ for different values of $\gamma_{1}$. 

When $\gamma = \Gamma$, we observe in the figure that the dynamics is similar to the one found in the Markovian case shown in Fig.~\ref{fig1markov} (a). However, at short times the characteristic non exponential behavior of the non-Markovian regime is present. In this case, a pointer state also emerges as evidenced by the frozen classical correlations. 

As the value of $\gamma$ decreases in relation to $\Gamma$, the non-Markovian behavior becomes more evident: For $\gamma = 0.1 \Gamma$, on one hand we observe that quantum discord decays asymptotically to zero and after a given time, a revival is observed, followed again by an asymptotic decay.  On the other hand, the classical correlations show similar behavior in Fig.~\ref{fig1markov} (a) and (b).  This points out an interesting regime where quantum correlations exhibit non-Markovian dynamics~\cite{Ple} while classical correlations  evolves within a Markovian frame. Now, when $\gamma = 0.01 \Gamma$  this mixed non-Markovian and Markovian behavior of quantum and classical correlations respectively is still present with the only difference that quantum correlations exhibits more revivals before disappears completely.

In the last case, $\gamma = 0.001 \Gamma$ the amplitude of the quantum discord revivals increases but its behaviors remains similar to previous cases. Interestingly, this is not true for classical correlations whose evolution experiences  significant differences with all previous cases (Markovian and non-Markovian) considered. Although the system settles on a pointer-state basis, this basis is no longer stable. Indeed, we observe the emergence of metastable pointer states as previously found in reference~\cite{Las}. 

\begin{figure}[t]
\includegraphics[width=80mm]{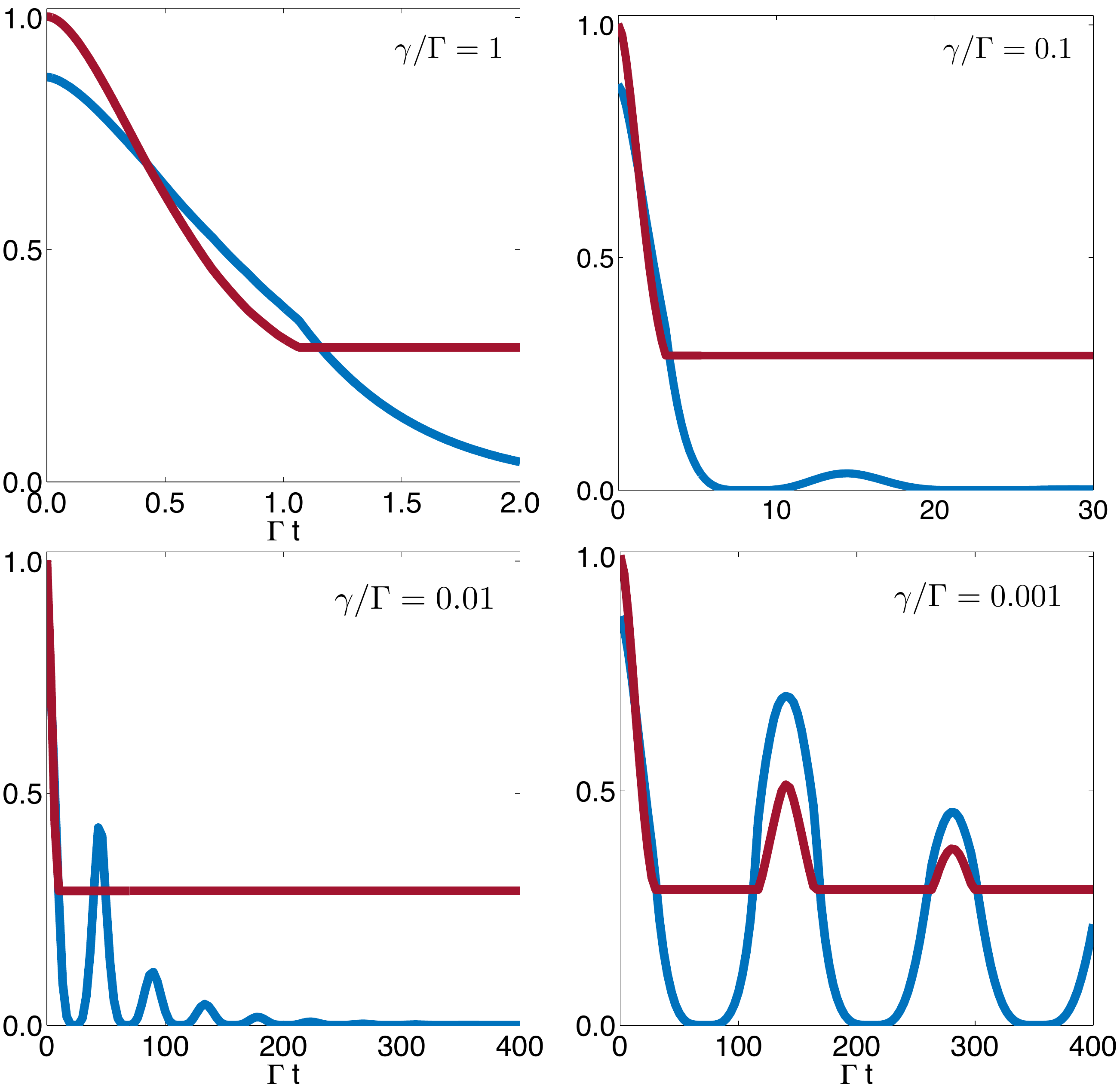} 
\caption{Evolution of  discord $D(\hat{\rho}_{AB})$ (blue) and classical correlations $C(\hat{\rho}_{AB})$ (red) for the initial state of Eq.~(\ref{eq:qubit-state})  with $(p_0=0.3; p_1 = 0.1; p_2=0.6)$ for different values of $\gamma/\Gamma $. }
\label{fignonmarkov} 
\end{figure}

\subsection{Long-Time limit}
The optimization process required to calculate the classical correlations defined in Eq.~(\ref{cc}), makes the search for analytical expressions of $C(\rho_{AB})$ a difficult task to realize in general. For instance, in qutrits 
this has been made numerically only~\cite{retamal17}. However, in our physical system of two non-interacting qutrits each under dephasing, the classical correlations can be calculated analytically when $t\rightarrow \infty$. After some calculations, we find that
 \begin{equation}
C(t \rightarrow \infty) = \max_{[\theta,\phi]} f(\theta,\phi),
\label{sol11}
\end{equation}
with
\begin{equation}
f(\theta ,\phi) = -\log_2 {(1/3)}+(1/3)\sum_{j=1}^{9}\lambda_{j}\log_2 (\lambda_{j}),
\label{f}
\end{equation}
\begin{figure}[t]
\includegraphics[width=70mm]{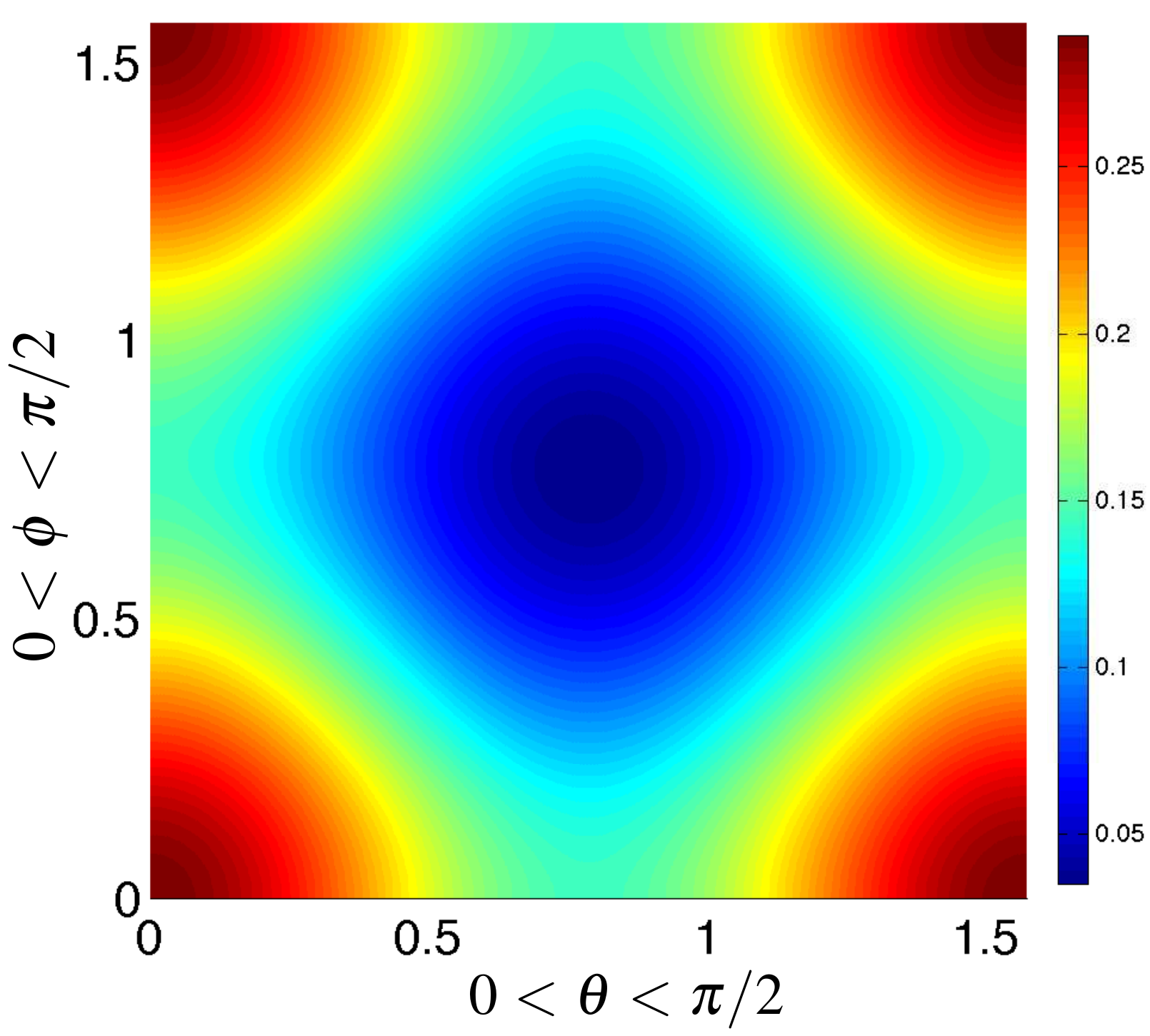} \caption{Function $f(p_0,p_1)$ using the normalization constrain $p_2=\sqrt{1-(p_0^2+p_1^2)}$, which for each value of $p_0,p_1$ we have performed the maximization over $\theta$ and $\phi$.
}
\label{function} 
\end{figure}
and 
\begin{eqnarray*}
\lambda_{1,2,3} &=&\left[x \cos^{2}(\theta)+\sin^{2}(\theta)[y \cos^{2}(\phi)+z \sin^{2}(\phi)]\right] \\
\lambda_{4,5,6}&=&\left[x \sin^{2}(\theta)+\cos^{2}(\theta)[y \cos^{2}(\phi)+z \sin^{2}(\phi)]\right] \\
\lambda_{7,8,9}&=&\frac{1}{2}\left[1-x+(y-z)\cos(2\phi)\right] 
\end{eqnarray*}
where $(x,y,z)$ is $(p_0,p_1,p_2)$, $(p_1,p_2,p_0)$ and $(p_2,p_0,p_1)$.

As we see from the expressions above, in the long-time limit $t \rightarrow \infty$, classical correlations depend only on the angles $\theta$ and $\phi$, rather than the four original parameters defined by the orthonormal base $\{|V_1\rangle,|V_2\rangle,|V_3\rangle\}$ previously defined. Fig.~\ref{function}, shows the function $f(\theta, \phi)$ from which the classical correlations  is calculated [Eq.~(\ref{sol11})]. It is clear in the Fig. that the maximum value of this function is found at four different sets of angles $(\theta, \phi)$. The four points corresponds to the following set of angles: $(\theta=0,\phi=0)$; $(\theta=\pi/2,\phi=0)$; $(\theta=0,\phi=\pi/2)$ and $(\theta=\pi/2,\phi=\pi/2)$. Therefore, using these results we can reconstruct the basis from which classical correlations are obtained. For instance, from the set $(\theta=0,\phi=0)$ we have   
\begin{eqnarray}
|V_{1}\rangle &=& |2\rangle \notag\\
|V_{2}\rangle &=& e^{i\chi_1}|0\rangle \label{base}\\
|V_{3}\rangle &=& e^{i\chi_2}|1\rangle. \notag
\end{eqnarray}
Notice that the angles $\chi_1$ and $\chi_2$ may take any value in the interval $0 \le \chi_1,\chi_2\le 2\pi$ without changing the value for the classical correlations.  This result allow us to calculate analytically the classical correlations in the long-time limit, as a function of the initial state parameters $(p_0,p_1,p_2)$ as shown in Fig.~\ref{p0p1p2}. The stationary behavior of classical correlations shown in Fig.~\ref{p0p1p2} are in agreement to the previous findings showed in Fig.~\ref{fig1markov}. For example, maximum stationary values of classical correlations are found either $p_0$, $p_1$ or $p_2$ are equal to 1, that is, a pure initial state as shown in Fig.~\ref{fig1markov}(c). On the other hand, minimum values for frozen classical correlations are observed for a balanced superposition in the initial state: $p_0=p_1=p_2=1/3$ whose dynamics is shown in Fig.~\ref{fig1markov}(d).

\begin{figure}[t]
\includegraphics[width=70mm]{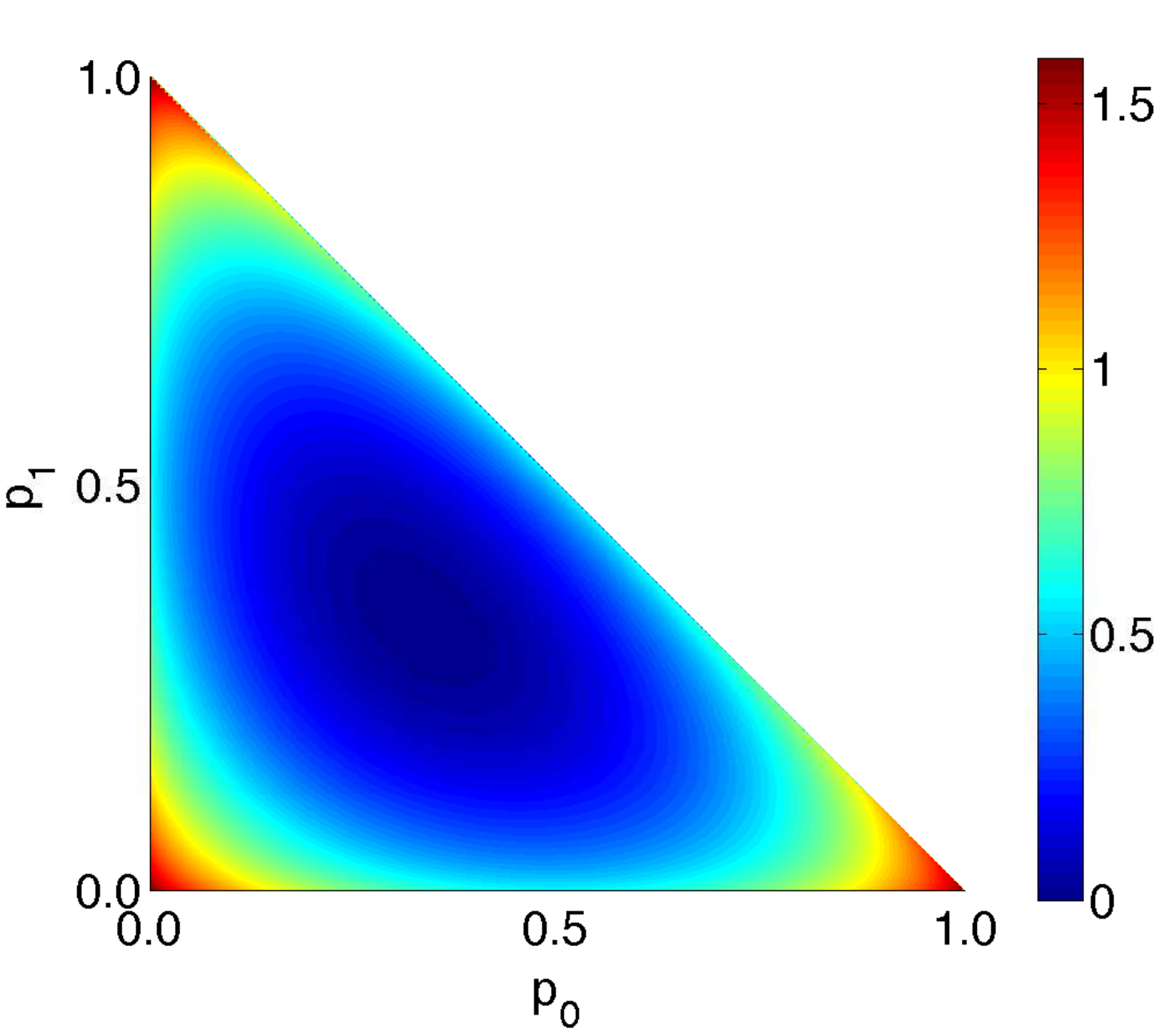} \caption{Stationary classical correlations $C(\hat{\rho}_{AB})$ in the long-time limit as function of $(p_0,p_1)$  }
\label{p0p1p2} 
\end{figure}

Although Markovian and non-Markovian regimes show important differences in the quantum dynamics of the system, as well as in the behavior of quantum and classical correlations, we found from our calculations that the settling of the system on the pointer state basis does not depend on the relation between the coherence times of the reservoir and the decay rate of the qutrits. In the Markovian regime, we see in Fig.~\ref{fig5}(a) the classical correlations calculated using the general form of Eq.~(\ref{cc}) together with the classical correlations obtained analytically. This figure shows us that the system settles on the pointer states basis found in the long-time limit,  long before this limit is actually reached. This is also the case when the non-Markovian regime is considered [Fig.~\ref{fig5}(b)], even though the settling might be unstable, at the end it reaches the same expected pointer state basis. Interestingly, the set of vectors defining the pointer states basis are the eigenvectors of the observable $\hat{S}_z$ with spin $s=1$. This analytical result is in agreement to what was observed in previous works that showed that eigenvectors of $\hat{S}_z$ with spin $s=1/2$ maximize classical correlations for qubits \cite{Cor,Las,Chen}.

\begin{figure}[t]
\includegraphics[width=90mm]{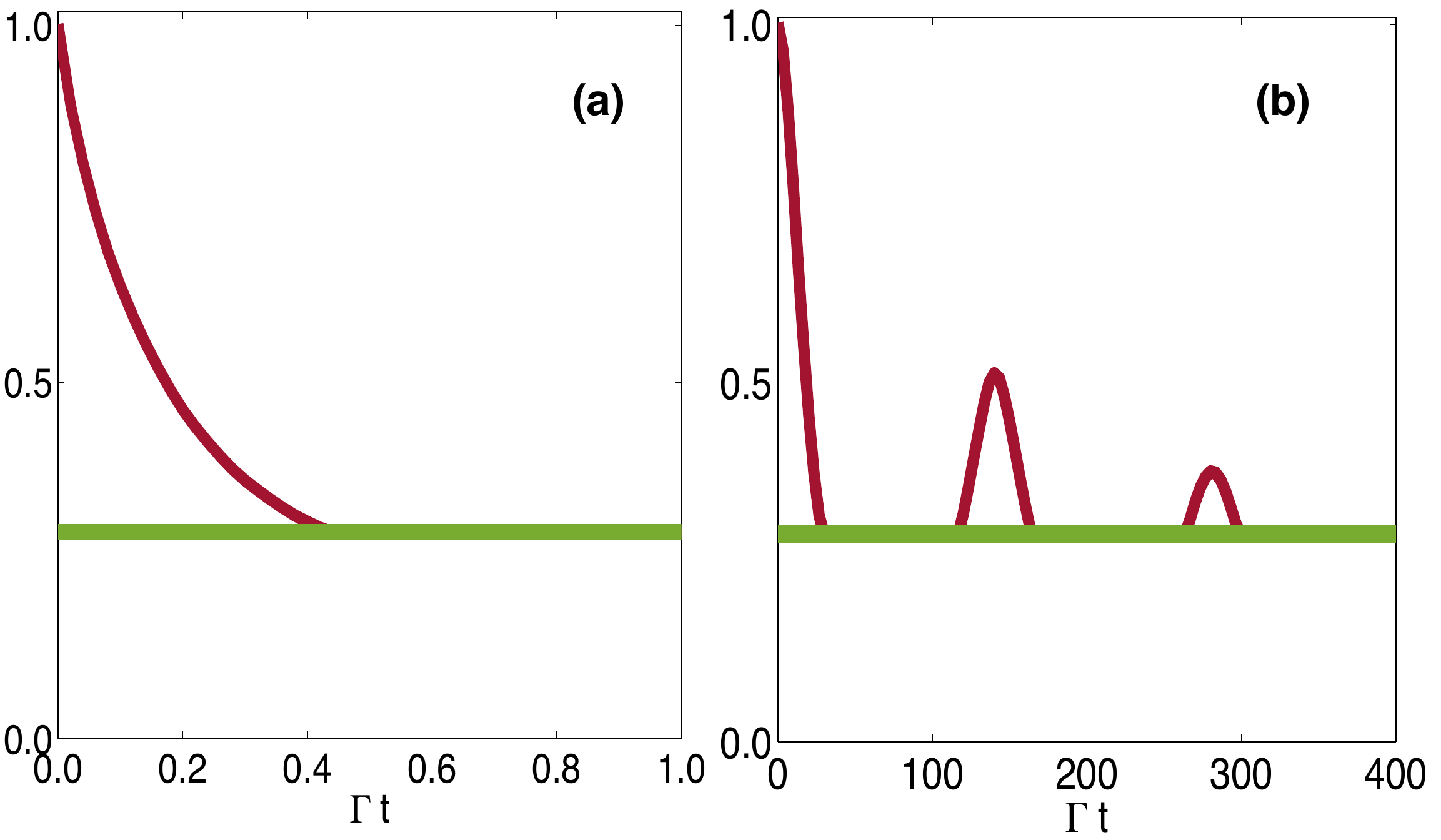} \caption{ Classical correlations calculated using the general form of Eq.~(\ref{cc}) (red) and considering the measurement on the second qutrit performed in the basis analytically found in Eq.~(\ref{base}) (green). (a) Markovian regime; (b) Non-Markovian case with $\gamma/\Gamma = 0.001$. Initial states corresponds to Eq.~(\ref{eq:qubit-state}) with $(p_0,p_1,p_2)=(0.3,0.6,0.1)$.}
\label{fig5} 
\end{figure}

\section{Concluding remarks}
In summary we have studied the dynamics of quantum and classical correlations in a two-qutrit system under the onset of dephasing. In the Markovian regime we observe that the decoherence process can be characterized considering two different regimes: first one where both classical and quantum correlations are affected by the decoherence, and a second regime where only quantum correlations decay while classical correlations remains constant. This can be understood as the measurement basis projecting the measured qutrit into classical states (pointer states) that are not affected by decoherence.  We show in the long-time limit that this observable corresponds to the spin operator $\hat{S}_z$ with spin $s=1$, consistent with previous similar results for the two-qubit case where the same operator but with spin $s=1/2$ is found. On the other hand, we find more varied results as a function of the ratio $\gamma/\Gamma$. For example, we find for $\gamma/\Gamma \ll 1$ that the non-Markovianity is reflected in the classical correlations by the apparition of metastables pointer states.

\section{ACKNOWLEDGMENTS}
C.E.L. acknowledges support by DICYT 041631LC.

\end{document}